# Three-dimensional Quantum Polarization Tomography of Macroscopic Bell States

Bhaskar Kanseri[1], Timur Iskhakov[1], Ivan Agafonov[2], Maria Chekhova[1,2], and Gerd Leuchs[1]

[1]*Max-Planck Institute for the Science of Light, G.-Scharowsky Str. 1/Bau 24, 91058, Erlangen, Germany*

[2]*M.V.Lomonosov Moscow State University, 119992 GSP-2, Moscow, Russia*

## ABSTRACT

The polarization properties of macroscopic Bell states are characterized using three-dimensional quantum polarization tomography. This method utilizes three-dimensional inverse Radon transform to reconstruct the polarization quasiprobability distribution function of a state from the probability distributions measured for various Stokes observables. The reconstructed 3D distributions obtained for the macroscopic Bell states are compared with those obtained for a coherent state with the same mean photon number. The results demonstrate squeezing in one or more Stokes observables.

PACS numbers: 42.25.Ja, 42.30.Wb, 03.65.Ud, 42.50.Lc

# I. INTRODUCTION

In recent years, macroscopic states of nonclassical light are a subject of intense research. Among many reasons, one important reason is the technological aspect as these states can provide much stronger interactions with matter and with each other than their microscopic counterparts [1-7]. One example of such macroscopic quantum systems is *macroscopic squeezed vacuum* [3]. An important advantage of macroscopic squeezed vacuum in comparison with conventionally used squeezed states (see, for instance, [8]) is that the former is fully nonclassical. Squeezed coherent states contain a huge component of classical (coherent) excitation and their squeezed vacuum part is rather weak, amounting to only few photons per mode. Various applications of macroscopic squeezed vacuum have been proposed, to name a few are macroscopic Bell tests [9], gravitational wave detection [10], quantum memory [11], absolute measurement of detectors' quantum efficiency [12] etc. Nonclassical states of light comprising polarization have been studied widely in the last couple of decades [13-20]. Some of such states, called polarization squeezed states, are characterized by the reduction of noise in specific polarization observables. A particular case of macroscopic squeezed vacuum involving two polarization modes and two frequency or wavevector modes, called *macroscopic Bell states,* is one such example [5, 14]. These multi photon states possess peculiar polarization properties. Despite the fact that all macroscopic Bell states are unpolarized in the first order of intensity, three of them (the triplet states) have suppression of noise in one of the three Stokes observables, whereas the fourth state (the singlet state, or *p-scalar light*) has suppression of noise in all the three Stokes observables.

Among several experimental tomography methods in quantum optics, the most widely studied and best implemented one is the field tomography of single-mode radiation [21-24]. This method relies on the tomographic representation of the density operator in the form of an integral expansion in some basis operators. The coefficients of this expansion are the probability distributions of the field observables (quadratures) obtained in homodyne measurements [23]. However, due to the inadequacy of single mode radiation model in optics, quantum tomography of multimode fields is required. In another method, called quantum tomography of polarization states, the polarization observables are functions of quadratures of several polarization modes [25]. However, it comprises experimental difficulties in implementation due to the use of multimode homodyne technique [26].

In this paper we aim to characterize macroscopic Bell states using a different approach known as *quantum polarization tomography*, in which the drawbacks of the former methods are overcome. This method was first proposed by Karassiov and Masalov and applied to unpolarized polarization-squeezed light generated at the output of the OPA [15, 16]. This method involves direct experimental reconstruction of polarization quasiprobability distribution (QPD) functions from probability distributions obtained in simple polarization measurements. In analogy with the field tomography, which requires recording of the probability distributions of a set of rotated quadratures, in polarization tomography we deal with a set of rotated Stokes polarization observables [15]. This method turns out to be a very efficient tool to probe the polarization properties of quantum states of light. A modification of this method, with the state reconstruction performed in separate spherical 'layers', was applied to the characterization of an intense polarization-squeezed state with a coherent polarized component [19]. The present work is the first attempt to characterize the polarization properties of the macroscopic Bell states using 3D quantum polarization tomography.

The brief outline of the paper is as follows. The polarization properties of light and their description using the Stokes observables in classical and quantum optics are summarized in Section 2. Section 3 comprises a brief introduction about the macroscopic Bell states. Here we consider two polarization and two frequency modes of radiation. In Section 4 the method of 3D quantum polarization tomography is described, with a brief mathematical background. In Section 5 we present the experimental method for the preparation of macroscopic Bell states and the method of measurement using a standard Stokes setup. Section 6 deals with the results of the reconstruction, their comparison and discussion.

## II. DESCRIPTION OF THE POLARIZATION IN TERMS OF THE STOKES OBSERVABLES

In general, polarized light is characterized using the Stokes observables. In classical description, the four Stokes parameters are introduced in terms of sums and differences of mean intensities in different polarization modes. Namely [27],

$$\begin{cases} S_0 = I_H + I_V \\ S_1 = I_H - I_V \end{cases}, \quad \begin{cases} S_2 = I_{H'} - I_{V'} \\ S_3 = I_+ - I_- \end{cases}; \qquad (1)$$

where $(H, V)$, $(H', V')$ and $(+, -)$ represent linear, $+45^0$ rotated linear and circular modes of polarization, respectively. It is evident that the $S_0$ parameter represents the total intensity whereas the remaining three parameters characterize the polarization properties of light. Conventionally, polarization states of light beams are classified as unpolarized, partially polarized and polarized. These states can be differentiated by the first-order degree of polarization, i.e., the length of the Stokes vector. The degree of polarization can be measured using the relation [13]

$$P = \frac{I_{max} - I_{min}}{I_{max} + I_{min}}, \qquad (2)$$

where $I_{max}$, $I_{min}$ are the maximum and the minimum intensities obtained by performing arbitrary polarization transformation and then passing the light through a polarizer. The polarization state of light is illustrated as a point on a unit-radius sphere called the Poincare sphere. However, this representation is not valid for unpolarized and partially polarized light beams. The three-dimensional Stokes space where the radius is not fixed can be used to represent the polarization state of such light beams [16].

In quantum description, the Stokes variables are observables and can be associated with Hermitian operators. Each of the three polarization Stokes operators is given by the difference of the photon numbers in two orthogonal polarization modes [13, 16],

$$\begin{cases} \hat{S}_0 = \hat{a}_H^\dagger \hat{a}_H + \hat{a}_V^\dagger \hat{a}_V \\ \hat{S}_1 = \hat{a}_H^\dagger \hat{a}_H - \hat{a}_V^\dagger \hat{a}_V \end{cases}, \quad \begin{cases} \hat{S}_2 = \hat{a}_{H'}^\dagger \hat{a}_{H'} - \hat{a}_{V'}^\dagger \hat{a}_{V'} \\ \hat{S}_3 = \hat{a}_+^\dagger \hat{a}_+ - \hat{a}_-^\dagger \hat{a}_- \end{cases}; \qquad (3)$$

where $\hat{a}_H^\dagger (\hat{a}_H)$, $\hat{a}_V^\dagger (\hat{a}_V)$ are photon creation (annihilation) operator in the horizontal and vertical polarization modes, respectively. The other polarization bases transformed from linear polarization basis $(H, V)$ are defined as

$$\begin{cases} \hat{a}_{H'} = (\hat{a}_H + \hat{a}_V)/\sqrt{2} \\ \hat{a}_{V'} = (-\hat{a}_H + \hat{a}_V)/\sqrt{2} \end{cases}, \quad \begin{cases} \hat{a}_+ = (\hat{a}_H - i\hat{a}_V)/\sqrt{2} \\ \hat{a}_- = (\hat{a}_H + i\hat{a}_V)/\sqrt{2} \end{cases}. \qquad (4)$$

The analysis of the fluctuations of the Stokes observables makes it possible to more appropriately classify the polarization states of light beams. For characterizing fluctuations, higher-order correlations are essential. For instance, the states broadly classified as unpolarized may manifest hidden polarization (light unpolarized in the first order of intensity but not for higher orders) [13, 16]. The degree of polarization for arbitrary orders can be introduced as [7]

$$P = \frac{(\Delta S_n^k)_{max} - (\Delta S_n^k)_{min}}{(\Delta S_n^k)_{max} + (\Delta S_n^k)_{min}}, \quad (5)$$

where $(\Delta S_n^k) = \langle (S_n - \langle S_n \rangle)^k \rangle$ is the $k^{th}$-order central moment of the corresponding Stokes variable. For the second-order degree of polarization k=2, and $\Delta S_n^2$ is the variance of the Stokes variable [13]. Alternatively, higher-order degree of polarization can be introduced in terms of higher-order correlation functions [13].

If fluctuations of the Stokes observables have to be taken into account, which is always the case in quantum optics, the polarization state cannot be depicted by a point on a sphere, like in the simplest classical description. It has to be represented by some three-dimensional object in the Stokes space, the position of which is given by the mean values of the Stokes observables and the sizes in different directions are determined by the corresponding fluctuations [14, 15, 19, 28]. The fluctuations (polarization noise) are especially important for unpolarized light, when the mean values of the Stokes observables vanish, showing no polarization structure of light. In this situation, if the fluctuations are different for different Stokes observables, such light is characterized by *hidden polarization* [15].

## III. MACROSCOPIC BELL STATES

A weakly pumped four-mode optical parametric amplifier can produce at its output, in addition to the vacuum, two-photon Bell states,

$$\begin{aligned}|\Psi^\pm\rangle &= \frac{1}{\sqrt{2}}(a_1^\dagger b_2^\dagger \pm b_1^\dagger a_2^\dagger)|\text{vac}\rangle, \\ |\Phi^\pm\rangle &= \frac{1}{\sqrt{2}}(a_1^\dagger a_2^\dagger \pm b_1^\dagger b_2^\dagger)|\text{vac}\rangle,\end{aligned} \quad (6)$$

where $a^\dagger, b^\dagger$ are photon creation operators in the horizontal and vertical polarization modes, respectively, and the subscripts 1, 2 denote frequency or wavevector modes. Here, we will consider frequency modes $\omega_1(\omega_2)$, but a similar consideration is valid for wavevector modes. At strong pumping, such an OPA generates not only two-photon states, but also higher-order Fock states. The states at its output can be written as [5]

$$\begin{aligned}\left|\Psi_{mac}^{\pm}\right\rangle &= e^{\Gamma\left(a_1^{\dagger}b_2^{\dagger} \pm b_1^{\dagger}a_2^{\dagger} + \text{H.C}\right)}\left|\text{vac}\right\rangle \\ \left|\Phi_{mac}^{\pm}\right\rangle &= e^{\Gamma\left(a_1^{\dagger}a_2^{\dagger} \pm b_1^{\dagger}b_2^{\dagger} + \text{H.C}\right)}\left|\text{vac}\right\rangle\end{aligned} \quad (7)$$

where $\Gamma$ is the parametric gain coefficient. Owing to their close resemblance with the two-photon Bell states [29], these states can be called macroscopic (many photon) Bell states. Moreover, the preparation schemes for both kinds of Bell states are similar [5, 30].

For macroscopic Bell states, the mean values of the polarization Stokes observables vanish, $\langle S_1 \rangle = \langle S_2 \rangle = \langle S_3 \rangle = 0$, showing that the states are unpolarized in the first order in the intensity. Thanks to this unpolarized behavior, the uncertainty relations $\Delta S_i \Delta S_j \geq |\langle S_k \rangle|$, $(i \neq j \neq k = 1, 2, 3)$ impose no restriction on the noise suppression in all the Stokes observables simultaneously [16]. Furthermore, the noise in one Stokes observable can be suppressed completely. Thus, in addition to the states having one polarization observable completely noiseless, a state with all Stokes observables having no noise can be obtained. For the triplet macroscopic Bell states, i.e., $\left|\Psi_{mac}^{+}\right\rangle, \left|\Phi_{mac}^{-}\right\rangle$, and $\left|\Phi_{mac}^{+}\right\rangle$, fluctuations are suppressed for $S_1$, $S_2$ and $S_3$, respectively. For these states, the degree of polarization in the second order does not vanish and thus, these states manifest hidden polarization. On the other hand, the singlet state $\left|\Psi_{mac}^{-}\right\rangle$ has noise suppressed in all Stokes observables simultaneously. Since this state is unpolarized in all orders of the intensity, it is sometimes referred to as polarization scalar (P-scalar) light [16]. Note that the principal difference of these states from ones with a classical polarized component is that the macroscopic Bell states "sit" at the origin of the Stokes space and the concept of the Poincare sphere is absolutely inapplicable to them.

## IV. QUANTUM POLARIZATION TOMOGRAPHY

Quasiprobability functions are among the most important instruments in quantum mechanics. One uses them for the description of states in terms of non-commuting sets of observables, like, for instance, coordinate and momentum. Although joint probability distributions do not exist for non-commuting variables, they can be substituted by quasi-probabilities, providing some information on the state but not necessarily satisfying formal requirements to probabilities [31-33]. In a nutshell, polarization quasiprobability function provides a way to calculate the mean values and higher-order moments of Stokes observables. Therefore, it enables the visualization of polarization squeezing. Out of many choices of such distributions, the convenient one is the polarization quasiprobability function introduced by Wolf and Atakishiyev [15, 34, 35]. The polarization quasiprobability function $W(S_1, S_2, S_3)$ of some radiation state is a function of three real variables corresponding to the Stokes observables. It is given by the three-dimensional Fourier transform of the quantum polarization characteristic function $\chi(u_1, u_2, u_3)$,

$$W(S_1, S_2, S_3) = \frac{1}{(2\pi)^3} \iiint \chi(u_1, u_2, u_3) e^{-i(u_1 S_1 + u_2 S_2 + u_3 S_3)} du_1 du_2 du_3, \tag{8}$$

where the characteristic function is defined as $\chi(u_1, u_2, u_3) = \left\langle e^{i(u_1 S_1 + u_2 S_2 + u_3 S_3)} \right\rangle$ (the angular brackets denote the averaging over the quantum state). Out of several QPD functions, the polarization quasiprobability function is of particular interest because it provides the quantitative polarization analysis of quantum radiation state in the form closest to the classical description and simultaneously retains the quantum distinctiveness [14].

Quantum polarization tomography is a method for the reproduction of polarization QPD function from the simple polarization measurement results that characterize the quantum state of an object [15, 16]. In other words, similar to the classical tomography where the image of an object is reconstructed using the projections taken for different observation directions, in quantum polarization tomography, we reconstruct the polarization QPD function using the probability distributions obtained by taking measurements along different directions in the Stokes space. To characterize the polarization properties of quantum states, quantum polarization tomography is advantageous over the field tomography as it does not require the homodyne technique for the measurement.

The reconstruction procedure proposed in Ref. [15] is similar to the classical 3D tomography. There the image of an object is reconstructed by integrating all the filtered planar projections taken along the different directions of the object (see Fig. 1). The filtered planar (back) projection is simply the second derivative of the original planar projection [36]. This transformation from the projection space to the object space can be written using the three-dimensional inverse Radon transform [37]

$$f(\vec{X}) = -\frac{1}{8\pi^2} \int_0^\pi \sin\vartheta d\vartheta \int_0^{2\pi} d\varphi \left(\frac{\partial^2 f(r,\vec{n})}{\partial r^2}\right)_{r=\vec{X}\cdot\vec{n}}, \tag{9}$$

where $\vec{X} = (x, y, z)$, $f(\vec{X})$ is a three-dimensional object and $f(r,\vec{n})$ is the projection of the object in the direction of a unit vector $\vec{n}$. The distance of any arbitrary plane from the origin is given by $r$, as shown in Fig. 1.

In a similar manner, in quantum polarization tomography, the polarization QPD is reconstructed by taking the second derivative of the probability distribution of the Stokes observables obtained for different directions in the Stokes space and then summing them up within one hemisphere [16]. It is emphasized that due to the inversion symmetry of the measurement, each direction of one hemisphere corresponds to some direction in another hemisphere. Therefore, to avoid the redundancy of the data, measurements in one hemisphere are sufficient to reconstruct the quantum state.

An arbitrary Stokes observable $(S, \vartheta, \varphi)$ corresponds to the operator defined in terms of the Stokes operators $\hat{\vec{S}} = \{\hat{S}_1, \hat{S}_2, \hat{S}_3\}$ and a unit vector on the Poincare sphere $\vec{n} \equiv \{\sin\vartheta\cos\varphi, \sin\vartheta\sin\varphi, \cos\vartheta\}$ with angular coordinates $\vartheta, \varphi$ on the unit radius sphere [14],

$$(\hat{S}, \vartheta, \varphi) = \hat{\vec{S}} \cdot \vec{n} \equiv \hat{S}_1 \sin\vartheta\cos\varphi + \hat{S}_2 \sin\vartheta\sin\varphi + \hat{S}_3 \cos\vartheta. \tag{10}$$

Similar to Eq. (9), the reconstruction yields

$$W(\vec{p}) \propto -\int_0^{\pi/2} \sin\vartheta d\vartheta \int_0^{2\pi} d\varphi \frac{d^2 H(S, \vartheta, \varphi)}{dS^2}\bigg|_{S=\vec{p}\cdot\vec{n}}, \tag{11}$$

where $\vec{p} \equiv r\{\sin\theta\cos\phi, \sin\theta\sin\phi, \cos\theta\}$ denotes an arbitrary vector in spherical coordinates $r, \theta, \phi$ and $H$ is the probability distribution corresponding to the Stokes observable.

For quantum polarization tomography, not only the mean values and variances of the Stokes observables but their probability distributions (histograms) $H(S, \vartheta, \varphi)$ are also required. These distributions are obtained from the best fitting of these histograms plotted for each of the measurements. In our case, all the histograms of the measurements were well approximated by Gaussian distributions. Thus, for all $\vartheta, \varphi$, we have obtained

$$H(S,\vartheta,\varphi) = e^{-(S-\langle S\rangle)^2/2(\Delta S)^2}, \tag{12}$$

where $\langle S\rangle$ and $\Delta S$ are the mean value and the standard deviation, respectively, different for all histograms. It is apparent from Eq. (12) that the parameters characterizing the shape of the polarization QPD function are the mean values of the Stokes operators and their noise. Using Eq. (11), and replacing the integration by summation over the hemisphere, we obtain the polarization QPD in spherical coordinates

$$W(r,\theta,\phi) \propto -\Delta\vartheta\Delta\varphi \sum_{i,j=1}^{N,M} \sin\vartheta_i H_s^{//}(r,\theta,\phi,\vartheta_i,\varphi_j), \tag{13}$$

where $\Delta\vartheta$ and $\Delta\varphi$ are the angular step sizes (in spherical coordinates) corresponding to the half-wave plate and the quarter-wave plate, respectively, rotated in a grid of $N\times M$ to cover one hemisphere of the Poincare sphere.

## V. EXPERIMENTAL

The preparation method of macroscopic Bell states relies on the frequency non-degenerate parametric down conversion [38] in two type-I BBO crystals with thickness 2 mm and the optic axes in orthogonal planes (see Fig. 2). The signal and idler wavelengths are 635 nm and 805 nm, respectively. The crystals are pumped by a Nd:YAG laser third harmonic ($\lambda_{pump}$=355 nm, repetition rate 1 kHz, pulse duration 18 ps, energy per pulse 0.2 mJ). The orthogonally polarized squeezed vacuums at the output of the crystals are superposed using a polarization beamsplitter (PBS). The pump is eliminated using a dichroic mirror (DM) and a long-pass filter (OG). The relative phase between the two squeezed vacuums can be varied with the help of a trombone prism (Fig. 2). If the phase is equal to zero, the superposition gives the macroscopic Bell state $|\Phi_{mac}^+\rangle$. With the relative phase equal to $\pi$, the resulting macroscopic Bell state is $|\Phi_{mac}^-\rangle$. In a $45^0$ rotated basis, the state $|\Phi_{mac}^-\rangle$ becomes $|\Psi_{mac}^+\rangle$. Using a dichroic plate (DP), which introduces a $\pi$ difference between the ordinary and extraordinary phase delays at the wavelengths 635 nm and 805 nm, the state $|\Psi_{mac}^+\rangle$ is converted into $|\Psi_{mac}^-\rangle$ [5, 30]. An aperture (A) was put at the focal plane of a lens (L) to select the angular spectra of the combined beam.

The measurement scheme is the standard Stokes measurement setup consisting of an achromatic half-wave plate (HWP) and a zero-order quarter-wave plate (QWP) followed by a Glan prism (GP). After the prism, the two orthogonally polarized output beams are detected by separate p-i-n diode detectors [3, 6]. The quantum efficiencies of the detectors for the wavelengths 635 nm and 805 nm are 85% and 95%, respectively. The output pulses from the detectors are measured using an analog-digital card, which integrates them over time. The resulting integrals, measured in units 'V·s' are linearly proportional to the photon numbers incident on the detectors during a light pulse.

For an orientation $\alpha$ of the half-wave plate and $\beta$ of the quarter-wave plate, the selected direction in the Stokes space is given in terms of the spherical coordinates by the transformations

$$\vartheta = \pi/2 - 2\beta,$$
$$\varphi = 2\beta - 4\alpha, \tag{14}$$

which define the unit vector $\vec{n}$ on the Poincare sphere.

In our experiment the half-wave plate and the quarter-wave plate were rotated in the steps of $2.5^0$ and $5^0$, respectively, each making 19 steps. The resulting orientations traced more than one quarter of the Poincare sphere (shown by points in Fig. 3 (a)). The points exceeding one quarter of the sphere were removed. The remaining points of the quarter sphere were reflected giving points covering exactly one hemisphere, as shown in Fig. 3 (b). The values of the Stokes observable for each direction of $\vec{n}$ were proportional to the difference of the signals obtained from the detectors. For each direction, 20000 pulses were measured giving a histogram for the Stokes observable.

## VI. RESULTS OF THE RECONSTRUCTION

### A. The triplet state $|\Phi^-_{mac}\rangle$

For the triplet states, the role of the polarization noise becomes very important, as these macroscopic Bell states manifest hidden polarization. For each of the macroscopic triplet states, fluctuations of a certain Stokes observable are suppressed. The triplet state prepared in our case is $|\Phi^-_{mac}\rangle$. For this state, the Stokes observable $S_2$ should not fluctuate.

Using the data acquired from the tomography measurement, the histogram for each measurement was fitted with a Gaussian distribution giving the mean value and the standard deviation (noise). The plots for the probability distributions and their Gaussian approximations for $S_1$ and $S_2$ measurement are shown in Fig. 4(a). It is evident from these plots that the noise for $S_2$ is smaller than the noise for $S_1$. These measurements comprise the contribution of electronic noise, which has to be deducted from the measured signal. Since the probability distributions for the noise and the signal are independent, the subtraction of the noise is given by the deconvolution of the measured signal and the noise. For Gaussian distributions, the deconvolution simply results in the subtraction of the electronic noise variance from the variance of the signal. After the deduction of the electronic noise, the mean and standard deviation values for each Stokes observable were put into Eq. (12) and the reconstructed quasiprobability distribution was plotted in three dimensions (Fig. 5(b)). The corresponding distribution without the electronic noise deduction can be seen in Fig. 5(a). We see that the reconstructed QPD function for the triplet state has an ellipsoidal shape confirming the suppression of noise in $S_2$ at the expense of noise enhancement in the other two Stokes observables. The electronic noise subtraction leads to a QPD function with more pronounced squeezing (see Fig. 5 (b)). Here and further after, all the reconstructed QPDs are normalized to the mean sum signal $S_0$ ($3\times10^{-6}$ V·s) which corresponds to approximately $3\times10^5$ photons [3]. Theoretically, the triplet state can be represented as a disk at the origin of the Stokes space having no noise for one (the squeezed) Stokes observable. However, experimentally there are many factors which restrict it from having no fluctuations. These are the non unity quantum efficiencies of the detectors, the optical losses and imperfections, the mismatch of the signal and the idler mode selection etc [6].

It is worth noting here that the QPD is a function of three variables. One method to visualize this function would be to plot all the points for which the value of the function is higher than some threshold. In our case, the threshold is $1/\sqrt{e}$ times the maximum value of the function. Thus the reconstructed object shows the $1/\sqrt{e}$ surface from the maximum of the QPD function.

## B. The singlet state

For the macroscopic singlet state $\left|\Psi^-_{mac}\right\rangle$, the degree of polarization is zero for all orders in the intensity. This state is unique in the sense that the variances for all the Stokes parameters are suppressed simultaneously. The plots of the probability distributions obtained for the observables $S_1$ and $S_2$, their Gaussian approximations and the probability distribution for the electronic noise are shown in Fig. 4(b). One can see that the noise is nearly the same for both Stokes measurements. The reconstructed QPD for the singlet state and for the electronic noise are shown in Fig. 5(c). It shows that the noise is equally suppressed in all the Stokes observables. After the electronic noise elimination, the reconstructed QPD is shown in Fig. 5 (d). It is a small sphere centered at the origin of the Stokes space, demonstrating fluctuations suppressed in all the Stokes observables simultaneously.

Theoretically, the singlet state can be represented as a point at the origin of the Stokes space having no noise for all Stokes observables. However, experimentally there are many factors which prevent it from having zero fluctuations. These are the non-unity quantum efficiencies of the detectors, the optical losses and imperfections, the mismatch of the signal and the idler mode selection etc [6].

## C. A coherent state

A coherent state can be defined as a boundary state between the classical and the nonclassical states. The variances of the Stokes observables for a coherent beam are all equal to the mean photon number of the beam (the shot noise limit). In general, this is the reason why a Stokes observable is said to be squeezed if its variance falls below the shot noise of a coherent beam having the same mean photon number [28].

To prepare a coherent state one needs a shot-noise limited source, i.e. a source for which the variance of the photon number scales linearly with the mean photon number. In general, for any source, the variance of photon number can be written as

$$Var(N) = \langle N \rangle + \left(g^{(2)} - 1\right)\langle N \rangle^2 , \qquad (13)$$

where $g^{(2)}$ is the second-order Glauber's correlation function [27]. The first term in the right-hand side denotes the shot noise whereas the second term describes the excess noise. For an ideal coherent source $g^{(2)} = 1$, and thus the excess noise vanishes. On the other hand, for a practical source $g^{(2)} > 1$ and the variance shows some quadratic dependence (see Fig. 6 (a)).

We used an intensity stabilized He-Ne laser as a source. The laser beam was reflected by a slit (≈150 μm) left on the blackened surface of a highly reflecting disk (a computer hard disk drive) spinning at 90 rotations per second. With this geometry, we obtained a source which mimicked a pulsed laser with a pulse width of ≈10 μs and a repetition rate of 90 Hz. The detectors were triggered by another pulse, of higher amplitude (obtained by making another slit of ≈1 mm size on the disk) which was separated from the first pulse by 750 μs. For this source, the dependence of the signal variance on the mean signal in one detector is shown in Fig. 6(a). The behavior is slightly non-linear, due to the unavoidable excess fluctuations. The sum signal for the detectors was taken to be the same as the one in the measurement of the squeezed states ($3 \times 10^{-6}$ Vs).

The reconstructed QPD function for the coherent state containing the electronic noise is shown in Fig 7 (a). The reconstruction object is displaced from the origin by the mean signal normalized to $S_0$ (proportional to the mean photon number). Due to the excess noise in the prepared state, the reconstruction results in a spheroid stretched in the $S_2$ direction, instead of a sphere. To compare the squeezed states with the coherent state, we displaced this spheroid to the origin and eliminated the electronic noise (see Fig. 7 (b)). The effect of the excess noise is much reduced in the case of balanced detection (when both detectors have the same signal), since in this case the excess fluctuations are cancelled out. For instance, as shown in Fig. 6 (b), for balance detection, the variance of the difference signal scales linearly with the mean sum signal. Therefore, to obtain the reconstruction of a coherent state (more precisely, in our case, it is a pseudo-coherent state), we took the standard deviation (noise) value that corresponded to the balanced detection for our source. The reconstructed QPD in this case (see Fig. 7 (b)) shows the same noise for all Stokes observables, which is equal to the shot noise of the source.

## D. Comparison of the reconstructed QPDs

The reconstructed QPD functions for the triplet and the singlet macroscopic Bell states, as well as for a coherent state with the same mean photon number, can be compared to observe the effect of squeezing. All these distributions are shown in Fig. 8. It is evident from the comparison between the polarization QPDs for the triplet and the coherent state that the triplet state has noise suppressed in observable $S_2$ whereas it is anti-squeezed in other Stokes observables. On the other hand, for the singlet state, noise for all Stokes observables is suppressed and is smaller than that for the coherent state (see Fig. 8).

It is worthwhile to mention here that, since the symmetries for the prepared quantum states were known, we reflected the mapped points obtained for the quarter sphere to cover the hemisphere, accordingly. However, for the tomography of any unknown quantum polarization state, the rotations of the half-wave plate and the quarter-wave plate should be chosen in such a way that at least one hemisphere of the Poincare sphere could be covered. Since this method relies on summing up (instead of integrating) over the hemisphere, the data points should have rather high density to approach the best results.

## VII. CONCLUSION

In conclusion, we have presented a reconstruction of macroscopic Bell states prepared via high-gain PDC in two type-I BBO crystals placed into a Mach-Zehnder interferometer. The reconstruction of polarization quasiprobability distribution functions from the polarization measurement results involves the method of 3D quantum polarization tomography. We observe that the polarization quasiprobability function, which serves as a quasi-classical portrait of the quantum polarization state of light, provides a more illustrative visualization of the polarization state of light than the Stokes observables. The resulting reconstructions for the triplet state and the singlet state were compared with the reconstructed PQD function of a coherent state showing squeezing in one and all Stokes observables, respectively. Not only these results illustrate the peculiar polarization properties of the polarization-frequency entangled states, but they also advocate the utilization of this direct reconstruction method for other quantum states. In future, these polarization rich states may found potential applications in testing the foundations of quantum theory e.g. Bell inequalities, separability, decoherence etc. in more involved manner.

## ACKNOWLEDGMENTS

We thank Falk Töppel for useful comments. T. Sh. I. acknowledges funding from Alexander von Humboldt Foundation.

**REFERENCES**

1. V. Vedral, Nature **453**, 1004 (2008).
2. F. De Martini, F. Sciarrino and C. Vitelli, Phys. Rev. Lett. **100**, 253601 (2008).
3. T. Sh. Iskhakov, M. V. Chekhova and G. Leuchs, Phy. Rev. Lett.**102**, 183602 (2009).
4. C. Vitelli, N. Spagnolo, L. Toffoli, F. Sciarrino and F. De Martini, Phys. Rev. A. **81**, 032123 (2010).
5. T. Sh. Iskhakov, M. V. Chekhova, G. O. Rytikov and G. Leuchs, Phys. Rev. Lett. **106**, 113602 (2011).
6. I. N. Agafonov, M. V. Chekhova, and G. Leuchs, Phys.Rev. A **82**, 011801 (2010).
7. T. Sh. Iskhakov, I. N. Agafonov, M. V. Chekhova, G. O. Rytikov and G. Leuchs, Phys. Rev. A **84**, 045804 (2011).
8. D. F. Walls and G. J. Milburn, *Quantum Optics* (Springer-Verlag, Berlin, 2008).
9. Ch. Simon and D. Bouwmeester, Phys. Rev. Lett. **91**, 053601 (2003).
10. K. McKenzie, D. A. Shaddock, D. E. McClelland, B. C. Buchler, and P. K. Lam, Phys. Rev. Lett. **88**, 231102 (2002).
11. L.V. Gerasimov, I.M. Sokolov, D.V. Kupriyanov, M.D. Havey, arXiv:1111.6669v1 [quant-ph] 29 Nov 2011.
12. I. N. Agafonov, M. V. Chekhova, T. Sh. Iskhakov, A. N. Penin, G. O. Rytikov, and O. A. Shcherbina, Opt. Lett. **36**, 1329, (2011).
13. D. N. Klyshko, Phys. Lett. A, **163**, 349 (1992); JETP **84**, 1065 (1997).
14. V. P. Karassiov, J. of Phys. A **26**, 4345 (1993); J. of Russ. Laser Res. **15**, 391 (1994); **21**, 370 (2000); **26**, 484 (2005).
15. P. A. Bushev, V. P. Karassiov, A. V. Masalov and A. A. Putilin, Opt. and Spectroscopy **91**, 526 (2001).
16. V. P. Karassiov and A. V. Masalov, Opt. Spectroscopy **74**, 551 (1993); Las. Phys. **12**, 948 (2002); J. Opt. B: Quant. and Semiclass. Opt. **4**, s366 (2002); JETP **99**, 51 (2004).
17. N. Korolkova, G. Leuchs, R. Loudon, T. C. Ralph, and Ch. Silberhorn, Phys. Rev. A, **65**, 052306 (2002).
18. R. Schnabel, W. P. Bowen, N. Treps, T. C. Ralph, H.-A. Bachor, and P. K. Lam, Phys. Rev. A **67**, 012316 (2003).

19. Ch. Marquardt, J. Heersink, R. Dong, M.V. Chekhova, A. B. Klimov, L. L. Sanchez-Soto, U. L. Andersen, and G. Leuchs, Phys. Rev. Lett. **99**, 220401 (2007).

20. A. Sehat, J. Söderholm, G. Björk, P. Espinoza, A. B. Klimov, and L. L. Sánchez-Soto, Phys. Rev. A **71**, 033818 (2005); L. L. Sánchez-Soto, E. C. Yustas, G. Björk, and A. B. Klimov, Phys. Rev. A **76**, 043820 (2007).

21. A. Royer, Found. Phys. **19**, 3 (1989).

22. K. Vogel and H. Risken, Phys. Rev. A **40**, 2847 (1989).

23. D. T. Smithey, M. Beck, M. G. Raymer and A. Faridani, Phys. Rev. Lett. **70**, 1244 (1993).

24. G. Breitenbach, S, Schiller and J. Mlynek, Nature **387**, 471 (1997).

25. H. Kuhn, D-G Welsch and W. Vogel, Phys. Rev. A **51**, 4240 (1995).

26. M. G. Raymer and M. Beck, *Experimental Quantum State Tomography of Optical Fields and Ultrafast Statistical Sampling*, Lect. Notes Phys. **649**, 235 (2004).

27. L. Mandel and E. Wolf, *Optical Coherence and Quantum Optics*, (Cambridge Univ. Press, New York, 1995).

28. W. P. Bowen, R. Schnabel, H. A. Bachor and P. K. Lam, Phys. Rev. Lett. **88**, 093601 (2002).

29. P.G. Kwiat, K. Mattle, H. Weinfurter, A. Zeilinger, A. V. Sergienko and Y. H. Shih, Phys. Rev. Lett. **75**, 433 (1995).

30. A. V. Burlakov, S.P. Kulik, G. O. Rytikov and M. V. Chekhova, JETP **95**, 639 (2002).

31. M. Hillery, R. F. O'Connell, M. O. Scully and E. P. Wigner, Phys. Rep. **106**, 121 (1984).

32. H-W. Lee, Phys. Rep. **259**, 147 (1995).

33. W. P. Schleich, *Quantum Optics in Phase Space*, (Wiley-VCH Verlag GmbH & Co. Berlin, Germany, 2005).

34. V. P. Karasev, Kratk. Soobshch. Fiz. Nos. **9-10**, 13 (1996); No. **9**, 34 (1999).

35. K. B. Wolf, Opt. Commun. **132**, 343 (1996); N. M. Atakishiyev, S. M. Chumakov, and K. B. Wolf, J. of Math. Phys. **39**, 6247 (1998).

36. M. Y. Chiu, H. Barrett and R. G. Simpson, J. Opt. Soc. Am. **70**, 755 (1980).

37. M. N. Wernick and J. N. Aarsvold, *Emission Tomography: The Fundamentals of PET and SPET*, (Elsevier Academic Press, California USA, 2004).

38. D. N. Klyshko, *Photons and Nonlinear Optics*, (Gordon and Breach, New York, 1988).

# FIGURE CAPTIONS:

**FIG.1.** A Planar projection in the 3D Stokes space. $\vec{n}$ is a unit vector. Integration over all planes orthogonal to the unit vector $\vec{n}$ gives the planar projection (tomogram) along this direction (shown as $f(r,\vec{n})$). The distance of any arbitrary plane from the origin is given by $r$. The sphere shows the coordinate axes corresponding to different Stokes observables.

**FIG.2. (a)** (Color online) Schematic of the experimental setup. The notation is described in the text. The states under study are produced from two non-degenerate collinear squeezed-vacuum beams generated via high-gain PDC in two type-I crystals placed into a Mach-Zehnder interferometer and pumped coherently. The registration part of the setup provides the measurement of various Stokes observables, depending on the orientations of the wave plates, and their probability distributions. The detectors give signals proportional to the detected photon numbers during a single light pulse. The difference signal of the two detectors corresponds to the measurement of a Stokes observable.

**FIG.3.** (Color online) Data points on the Poincare sphere corresponding to different orientations of the phase plates. **(a)** All directions resulting from the rotation of λ/2 (0-45$^0$) and λ/4 (0-90$^0$) in 19 steps each. **(b)** After the reduction to a quarter sphere and then reflection, the data points cover one complete hemisphere of the Poincare sphere.

**FIG.4.** (Color online) Probability distributions of the difference signal for **(a)** the triplet state $\left|\Phi^-_{mac}\right\rangle$, and **(b)** the singlet state. Squares represent the measurement for $S_1$ ($\vartheta = 90^0, \varphi = 0^0$) and triangles for $S_2$ ($\vartheta = 90^0, \varphi = 90^0$). In both plots, the probability distribution for the electronic noise is shown by circles. Lines show the Gaussian fits.

**FIG.5.** (Color online) The reconstructed polarization QPD functions (i) for the triplet state $\left|\Phi^-_{mac}\right\rangle$ (shown in blue color) **(a)** containing the electronic noise, and **(b)** after the subtraction of

the electronic noise; (ii) for the singlet state $\left|\Psi_{mac}^{-}\right\rangle$ (shown in red color) (**c**) containing the electronic noise, and (**d**) after the subtraction of the electronic noise. The QPD functions for the electronic noise are shown by black color in (**a**) and (**c**). Each of the plots demonstrates the surface on which the QPD function takes the value of $1/\sqrt{e}$ from its maximum. The mean values of the signals are normalized to $S_0$ whereas the standard deviation values are multiplied by 10 for better visualization.

**FIG.6.** (**a**) Dependence of the signal variance on the mean signal of one of the detectors for our laser source. The continuous line is the quadratic fit, $a+bx+cx^2$ ($a=3\times10^{-17}$ V$^2$s$^2$, $b=1\times10^{-11}$ Vs, $c=6.0\times10^{-6}$), showing some excess fluctuations. (**b**) Dependence of the variance of the difference signal on the mean sum signal for the case of balanced detection. The continuous line shows a linear fit.

**FIG.7.** (Color online) The reconstructed polarization quasiprobability distributions for a pseudo-coherent state (linearly polarized light in our case) after the subtraction of the electronic noise. (**a**) The state is displaced in the Stokes observable $S_1$ from the origin by the mean signal normalized to $S_0$ (linearly related to the mean photon number). (**b**) For the comparison with the triplet and the singlet states, the QPD is shifted to the origin (shown as a green oblate spheroid). In another case, the reconstruction was done with the standard deviation (noise) for the balanced detection in order to minimize the effect of excess fluctuations. This gives a sphere at the origin (shown in yellow color). For the better visualization, the standard deviation values are multiplied by 10.

**FIG.8.** (Color online) Comparison between the reconstructed polarization quasiprobability distributions of (1) a coherent state (yellow) (2) the $\left|\Phi_{mac}^{-}\right\rangle$ triplet state (blue) and (3) the singlet state (red).

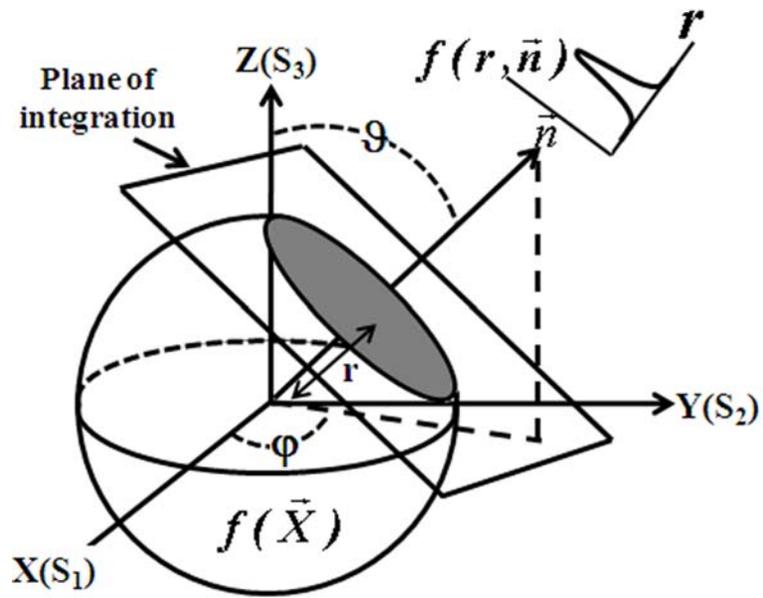

**FIG. 1**

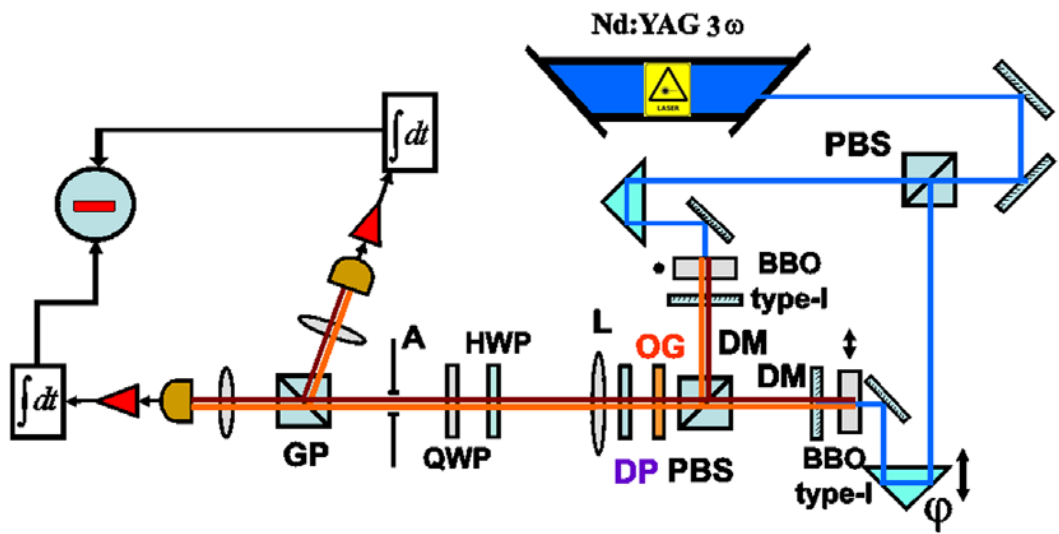

**FIG. 2**

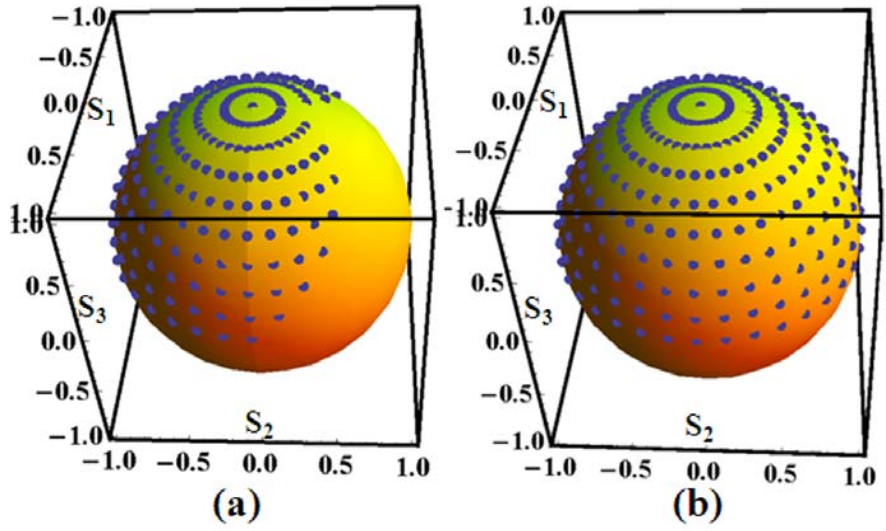

**FIG. 3**

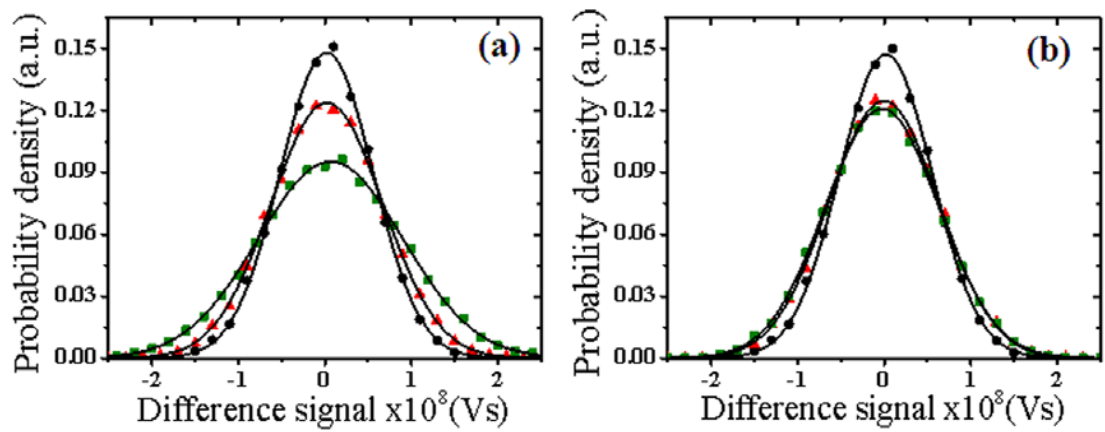

**FIG. 4**

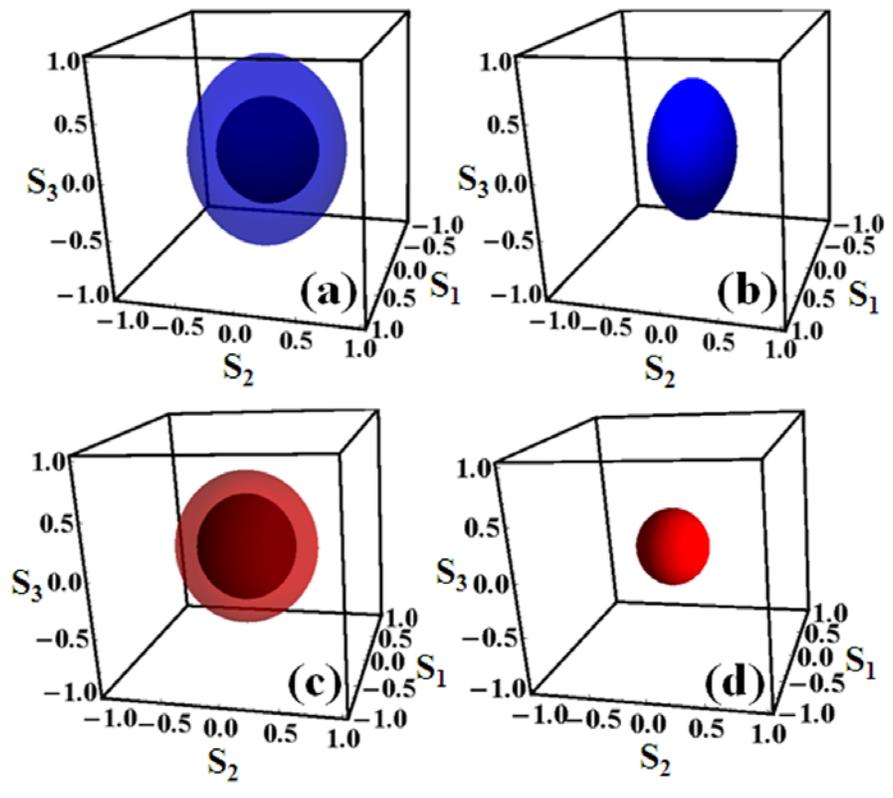

**FIG. 5**

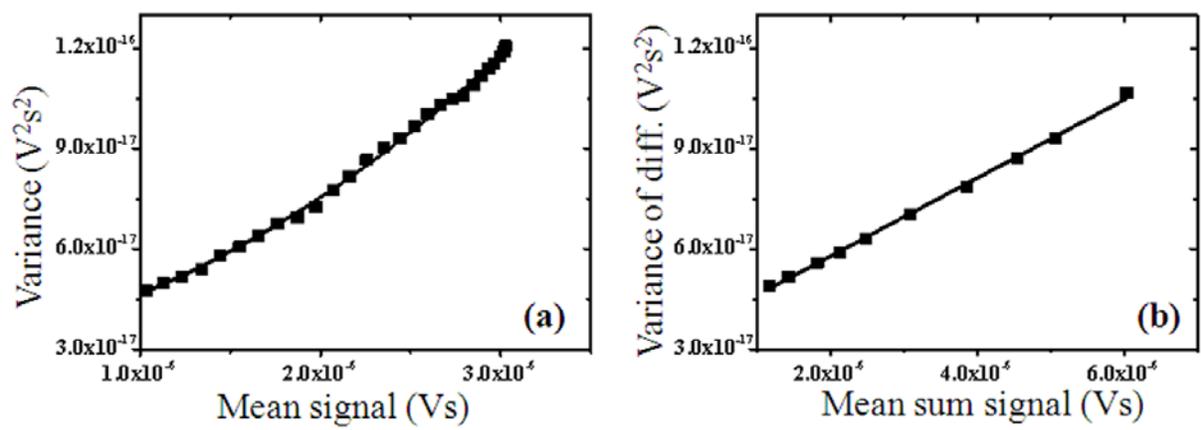

**FIG. 6**

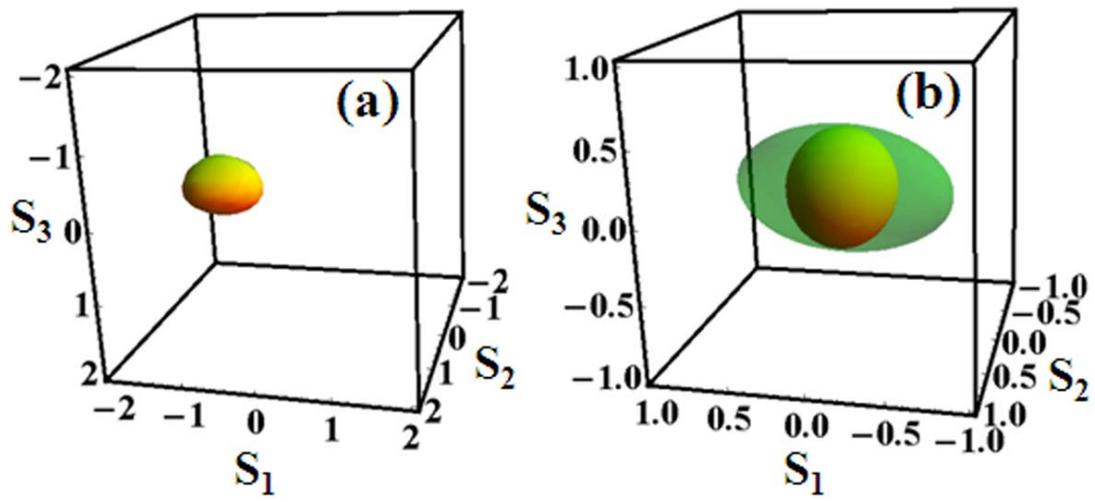

**FIG. 7**

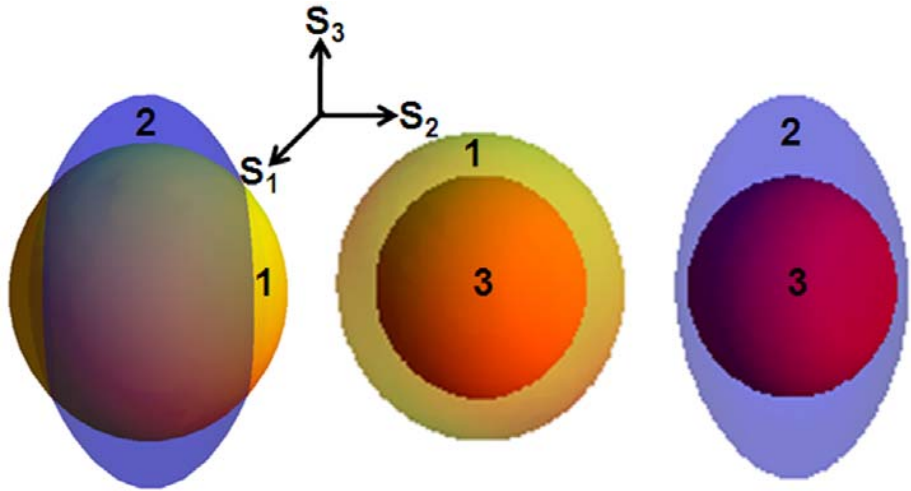

**FIG. 8**